\documentclass[showpacs,prl,twocolumn]{revtex4}
\usepackage{graphicx}

\begin{document}
\title[Electron Spin Resonance of YbRh$_{2}$Si$_{2}$]{Spin dynamics of \boldmath{$\mathrm{YbRh_2Si_2}$} observed by Electron Spin Resonance}
\author{J. Sichelschmidt$^1$, J. Wykhoff$^1$, H.-A. Krug von Nidda$^2$, J. Ferstl$^1$, C. Geibel$^1$ and F. Steglich$^1$}
\affiliation{$^1$Max Planck Institute for Chemical Physics of Solids, 01187 Dresden, Germany\\
                $^2$Experimentalphysik V, EKM, Universit\"at Augsburg, 86135 Augsburg, Germany}
\date{\today}

\begin{abstract}
Below the Kondo temperature $T_{\rm K}$ electron spin resonance (ESR) usually is not observable from the Kondo-ion itself because the characteristic spin fluctuation energy results in a huge width of the ESR line. The heavy fermion metal YbRh$_{2}$Si$_{2}$ seems to be an exceptional case where definite ESR spectra show characteristic properties of the Kondo-ion Yb$^{3+}$ well \textit{below} $T_{\rm K}$. We found that the spin dynamics of YbRh$_{2}$Si$_{2}$, as determined by its ESR relaxation, is spatially characterized by an anisotropy of the zero temperature residual relaxation only.
\end{abstract}


\maketitle
\section{Introduction}
The heavy-fermion compound YbRh$_{2}$Si$_{2}$ is located very close to the quantum critical point (QCP) corresponding to the disappearance of antiferromagnetic (AF) order (due to the increasing $f$-hybridization). Very weak AF order is observed at ambient pressure at $T_{\rm N}\sim70$~mK above which pronounced deviations from the Landau Fermi liquid behavior occurs. A highly anisotropic magnetic response indicates that Yb$^{3+}$-moments are forming an easy-plane square lattice perpendicular to the crystallographic $c$-direction \cite{tro00,geg02}. A tiny magnetic field of $B_{\rm c}=0.06$~T is sufficient to suppress the weak AF order.\\ 
\noindent The $T/(B-B_c)$ scaling observed in the low temperature specific heat and electrical resistivity points towards a locally critical (LC) scenario \cite{si01} for the QCP in this system \cite{cus03}. This scenario describes the break-up of heavy electrons at a QCP into a charge carrying part and a local spin part. The observation of an electron spin resonance (ESR) signal in YbRh$_{2}$Si$_{2}$ shows properties which unambiguously could be ascribed to the Kondo-ion Yb$^{3+}$ \cite{sic03}. Therefore this ''Kondo-ion ESR'' is a strong experimental evidence for the LC scenario and provides a unique opportunity for studying the dynamics of the local part. 
The observed ESR linewidth is about three orders of magnitude smaller than the linewidth $k_{\rm B}T_0 /g\mu _{\rm B} \approx 10$~T estimated from the spin fluctuation temperature $T_0\simeq24$~K inferred from thermodynamic measurements \cite{tro00}. In order to reveal the origin of the small linewidth additional information is desirable and has been found by investigating the ESR in YbRh$_{2}$Si$_{2}$ doped by Ge on the Si site \cite{sic05} or La on the Yb site \cite{wyk06}. Both dopands change the 4$f$-conduction electron hybridization and hence the distance to the QCP. The ESR properties seem to be sensitive to this distance as they can be described by a characteristic local moment screening temperature, $\tilde{T}_K$ \cite{sic03,sic05}. In this paper we focus on a more detailed analysis on the intensity, $g$--factor and the spatial dependence of the spin dynamics characterized by the ESR relaxation rate.
\section{Experimental Procedure and Results}
ESR probes the absorbed power $P$ of a transversal magnetic microwave field as a function of an external magnetic field $\vec{B}$. To improve the signal-to-noise ratio, a lock-in technique is used by modulating the static field, which yields the derivative of the resonance signal d$P$/d$B$. The ESR 
experiments were performed at X-band frequencies ($\nu $=9.4 GHz) and Q-band frequencies ($\nu $=33.1 GHz). The sample temperature was set with 
a $^{4}$He-flow-type ($T>4$~K) or a $^{4}$He-bath-type ($T>1.6$~K) cryostat, respectively. We used single crystalline platelets of YbRh$_{2}$Si$_{2}$ the preparation of which is described elsewhere \cite{tro00}. A large residual resistivity ratio of 68, a residual resistivity of less than 1 $\mu \Omega $cm 
as well as a very sharp anomaly in the specific heat at $T=T_{N}$ evidence the high quality of the single crystals. The concentration of Yb defect phases is less than 1{\%}. The magnetic and transport properties of the crystals have been thoroughly characterized \cite{tro00,geg02,cus03}.
\begin{figure}[h]
 \centering
 \includegraphics[width=0.47\textwidth]{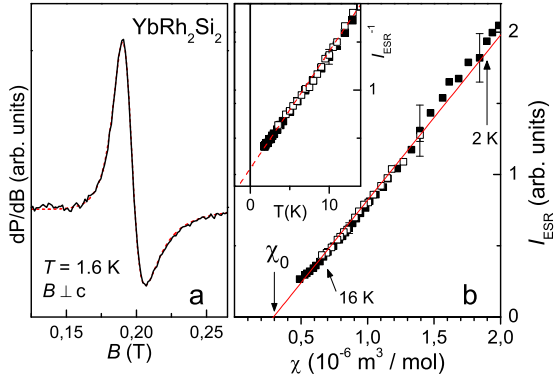}
\caption{a: Representative ESR spectrum at $T=1.6$~K. The dashed line fits the spectrum with a ``Dysonian'' shape. b: Basal plane ESR intensity, obeying a Curie-Weiss law (see inset), vs. basal plane static susceptibility $\chi $ at magnetic fields 0.19~T (X-band, closed symbols) and 0.68~T (Q-band, open symbols). The solid line extrapolates the data linearly and crosses zero intensity at $\chi_{0}$ which gives an estimate of a small temperature independent contribution in $\chi$. \label{Fig1}}
\end{figure}
\subsection{Data analysis of the ESR absorption line}
Figure \ref{Fig1}a shows a representative ESR signal at the lowest accessible temperature and with the external magnetic field $\vec{B}$ applied perpendicular to the crystallographic $c$-axis. The spectra could be fitted nicely with a Dysonian line shape \cite{feh55} providing the ESR parameters linewidth $\Delta B$, $g$-factor $(g=h\nu/\mu_{\rm B}B_{\rm res}$; $B_{\rm res}$: resonance field), and intensity. The observed asymmetry of the line is due to a non-vanishing ratio of dispersion to absorption ($D/A)$ contributions to the line and reflects a penetration depth being considerably smaller than the sample size \cite{feh55}. We used for all spectra $D/A$=1 without reducing the fit quality significantly among the spectra. We estimated that more than 60{\%} of the Yb$^{3+ }$ions in YbRh$_{2}$Si$_{2}$ must contribute to the observed ESR signal \cite{sic04}.
\subsection{Intensity}
The ESR intensity reflects the uniform static susceptibility of the ESR probe spin and is proportional to the area under the absorption 
signal. The temperature dependence of the penetration depth $\left[{\sigma (T)\cdot \nu } \right]^{-0.5} $ ($\sigma $: 
electrical conductivity) has been taken into account in the intensity data. The inset of Fig.~\ref{Fig1}b shows the basal plane ESR intensity, i.e. $\vec{B}\perp c$--axis. The temperature dependence follows nicely a Curie--Weiss law \cite{sic04} which demonstrates the local moment character of the  ESR probe spin. The very small Weiss temperature of $\Theta_{\bot}^{\rm ESR}=-1.5$~K approximately corresponds to the value which characterizes the Curie--Weiss type behavior of the bulk susceptibility for $T<0.5$~K \cite{geg03}. In order to illustrate the direct correspondence between the ESR intensity and the uniform static susceptibility Fig.~\ref{Fig1}b shows the ESR intensity plotted versus the susceptibility $\chi (T)$ which is measured at magnetic fields of 0.19~T (X-band resonance field) and 0.68~T (Q-band resonance field) applied in the tetragonal basal plane. A linear extrapolation of the ESR intensity crosses the zero intensity axis which estimates a small temperature independent contribution $\chi _0$ to the bulk susceptibility. As 
has been pointed out previously \cite{sic03} the obtained value $\chi _0 =\left({0.3\pm 0.1} \right)\times 10^{-6}{\rm m^3/molK}^2$ results in an enhanced Sommerfeld-Wilson ratio $R\cong 6$ which points towards the existence of pronounced ferromagnetic fluctuations that have been observed in NMR measurements \cite{ish02}.
\subsection{Anisotropy of the ESR $g$-factor}
\begin{figure}
 \centering
 \includegraphics[width=0.47\textwidth]{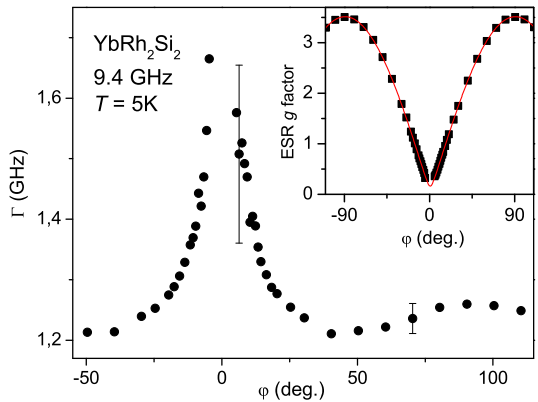}
\caption{Angular dependence of the ESR relaxation rate $\Gamma =g\mu _{B}\Delta B$/h at $T $= 5 K with the tetragonal $c$-axis enclosing the angle $\varphi$ with the external magnetic field and the microwave magnetic field applied within the tetragonal basal plane. Inset: $g(\varphi)$ at $T=5$~K showing uniaxial symmetry 
behavior with $g_\parallel$=0.17 and $g_{\bot }$=3.561 (solid line). \label{Fig2}}
\end{figure}

We identify the Yb$^{3+}$ spin with a $^{2}$F$_{7/2}$ ground state as origin for the anisotropy of the ESR in YbRh$_{2}$Si$_{2}$. Yb$^{3+}$ is located in an environment of uniaxial symmetry (point group $D_{4h})$ which results in strongly anisotropic magnetic properties, as seen in magnetic susceptibility measurements \cite{tro00}. This anisotropy is imposingly reflected in the angular dependence of the $g$-factor, see the inset of Fig.~\ref{Fig2}. A uniaxial anisotropy of the $g$-factor, $g(\varphi )=\sqrt {g_\parallel^2 \cos ^2\varphi +g_\bot^2 \sin^2\varphi } $ ($\varphi $: angle between the tetragonal $c$-axis and the external field $\vec{B}$) describes the data at $T=5$~K as shown by the solid line in the inset of Fig. \ref{Fig2}. However, within experimental accuracy, we found no anisotropy of the $g$-factor within the tetragonal basal plane of the crystal. That is, when rotating the crystal around the $c$-axis and keeping $\varphi=90^\circ$ the $g$-factor is independent on crystal orientation. Using the experimental $g$-values we can analyze the wave function of the two lowest Kramers doublets of the Yb$^{3+}$ ground state $^{2}$F$_{7/2}$. In a tetragonal environment the wavefunctions with symmetries $\Gamma _{6}$ or $\Gamma _{7}$ are \cite{kur65}:
\begin{eqnarray}
\label{eq1}
\Gamma_6&:& \sin\Theta_6\mid\pm1/2\rangle+\cos\Theta_6\mid\mp7/2\rangle \\
\Gamma_7&:& \sin\Theta_7\mid\pm5/2\rangle+\cos\Theta_7\mid\mp3/2\rangle  \nonumber
\end{eqnarray}
The corresponding $g$-values are ($g_{J} =8/7$):
\begin{eqnarray}
\label{eq2}
\Gamma_6 : &g_\parallel =& g_J \mid \sin^2\Theta_6 - 7 \cos^2\Theta_6\mid \\
                     &g_\perp =& 4g_J\sin^2\Theta_6 \nonumber \\[0.2cm]
\Gamma_7 : &g_\parallel =& g_J \mid 5\sin^2\Theta_7 - 3 \cos^2\Theta_7\mid \nonumber \\ 
                     &g_\perp =& 4g_J\sqrt{3}\mid\sin\Theta_7\cos\Theta_7\mid \nonumber
\end{eqnarray}
In order to reproduce the wavefunctions of eq.\ref{eq1} satisfactorily by the experimentally found $g$-values one needs to take into account that in metals the observed $g$-factor is shifted respective to its value in an ionic, insulating environment $\Delta g=g_{\rm metal}-g_{\rm ionic}$. For the isothermal case $\Delta g \propto N(\epsilon_{\rm F})J(q=0)$ where $N(\epsilon_{\rm F})$ is the density of states at the Fermi energy and $J(q=0)$ is the local moment -- conduction electron exchange integral at wavevector $q=0$ \cite{tay75}. In order to reconcile the large $g$--factor anisotropy ($g_\perp/g_\parallel\simeq21$) with the wavefunction symmetries $\Gamma_6$ or $\Gamma_7$  $\Delta g$ should be negative.  The Yb$^{3+}$ resonance of the non-metal tetragonal system PbMoO$_{4}$ yields $g_{ionic}^{\bot }$ =3.86 \cite{kur65}, i.e. a relative $g$-shift may be up to $-10\%$. Thus, for $g_\parallel=0.17\pm0.07$ the absolute shift would be within the error if an isotropic $g$--shift is assumed. We therefore insert $g_\parallel$ in equations (\ref{eq2}) and get $g_{\bot }^{a}(\Gamma _{6})$ = 3.92$\pm $0.04 , $g_{\bot}^{b}(\Gamma _{6})$ = 4.08$\pm $0.04 and $g_{\bot }^{a}(\Gamma_{7})$ = 3.80$\pm $0.03, $g_{\bot }^{b}(\Gamma _{7})$ = 3.86$\pm$0.03, where the indices a,b denote two possible solutions. If we assume $\Delta g\simeq-10${\%} these $g_{\bot }$-values match the measured ones and therefore the angles in equations (\ref{eq1}) can be restricted to $67^\circ\le\Theta _{6 }\le72^\circ$ and $36^\circ\le\Theta _{7}\le39.5^\circ$. The $\Gamma _{7}$ ground state symmetry appears to be more likely, as the corresponding $g_{\bot }$ values are closer to the measured ones.
\subsection{Anisotropy of the relaxation}
In order to characterize the spatial dependence of the spin dynamics of the ESR probe spin we investigated the angular dependence of the ESR relaxation rate $\Gamma=g\mu_{\rm B}\Delta B/h$ in the temperature range 4.2--12~K. The dissipation-fluctuation theorem relates $\Gamma $ to the imaginary part of the dynamic susceptibility, \mbox{$\Gamma\propto T/\omega\cdot{\rm Im}\chi(\bar{q},\omega)$}, where $\bar{q}$ denotes a $q$-space average of $\chi$ and $\omega=2\pi\nu$. The angular variation of the relaxation rate $\Gamma(\varphi)=g(\varphi)\mu_{\rm B}\Delta B(\varphi)/h$ is shown in Fig.~\ref{Fig2} at $T=5$~K. The tetragonal $c$-axis encloses the angle $\varphi$ with $\vec{B}$ and the microwave magnetic field $\vec{b_0}$ is always perpendicular to $c$. Within the experimental accuracy a pronounced deviation from the $\Gamma(\varphi=90^\circ)$--value is visible for $\varphi\le30^\circ$. However, when keeping $\varphi$ fixed we found the relaxation rate to be independent on how the crystalline axes are oriented in $\vec{b_{0}}$. 

\noindent For each angle the temperature dependence of the linewidth can be well described by \mbox{$\Delta B(T)=\Delta B_0+\beta\cdot T+\frac{c\cdot \Delta}{\exp(\Delta/T-1)}$} \cite{sic03}. The exponential part dominates above $T \cong 12$~K and describes the relaxation via conduction electrons incorporating the first excited crystal field level at $\Delta\cong10$~meV \cite{hir69}. The slope $\beta$ of the $T$-linear part of $\Delta B(T)$ corresponds to a slope $\partial\Gamma/\partial T$ in the $T$-linear part of the relaxation rate $\Gamma(T)=g(T)\mu_{\rm B}\Delta B(T)/h$.
\begin{figure}
 \centering
 \includegraphics[width=0.47\textwidth]{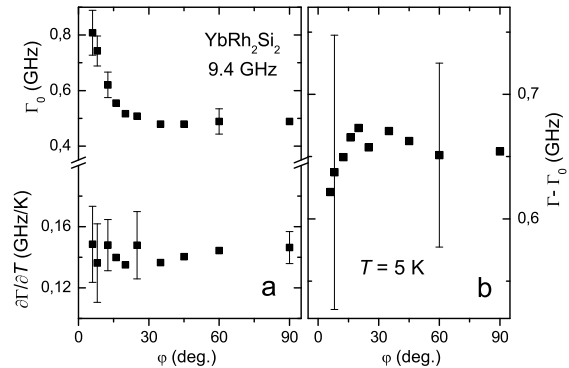}
\caption{a: Angular dependence of the residual relaxation rate $\Gamma _0=\Gamma(T=0\,{\rm K})$ and the slope $\partial\Gamma/\partial T$ of 
the $T$--linear part of $\Gamma(T)$. b: Angular dependence of the ESR relaxation corrected by the residual relaxation 
(rotation geometry as in Fig.~\ref{Fig2}).\label{Fig3}}
\end{figure}

\noindent The angular variation of $\partial\Gamma/\partial T$ and of the residual relaxation $\Gamma _{0}$ is shown in Fig.~\ref{Fig3}a.
$\partial\Gamma/\partial T(\varphi)$ appears \textit{isotropic} within the error bars. In the isothermal case within a Fermi liquid theory $\partial\Gamma/\partial T\propto \left[ N(\epsilon_{\rm F})J \right]^2$ \cite{tay75}. The zero temperature residual relaxation rate $\Gamma _{0 }$ shows a pronounced anisotropy at $\varphi\le30^\circ$. A very similar behavior is found in the total relaxation rate $\Gamma $, see Fig.~\ref{Fig2}. Therefore, as shown in Fig.~\ref{Fig3}b, we claim that $\Gamma-\Gamma _{0}$ behaves spatially isotropic within our experimental accuracy.
\section{Conclusion}
We conclude that the large magneto-crystalline anisotropy \cite{tro00} of YbRh$_{2}$Si$_{2}$ is clearly reflected in a large ESR $g$--factor anisotropy, which may be reasonably described by a $\Gamma _{7}$ ground state symmetry of the Yb$^{3+}$ ion. The spatial variation of the spin dynamics, as seen by the ESR relaxation rate, shows anisotropy only in the zero temperature residual relaxation rate. The isotropic behavior at finite temperatures of the dynamic spin properties of the Kondo-ion is consistent with a $q$ wavevector independent form of the spin susceptibility within the LC scenario of quantum criticality \cite{si01}. The temperature dependence of the ESR intensity indicates a contribution of ferromagnetic ($q=0$) fluctuations at the resonance magnetic fields, 0.19~T and 0.68~T.
\section{Acknowledgements} 
The low-temperature ESR-measurements at the University of Augsburg were partially supported by the German BMBF under Contract No. VDI/EKM13N6917 and by the Deutsche Forschungsgemeinschaft within SFB484 (Augsburg).
%
%
%

\end{document}